# The best whistler: a cavitating tip vortex

Zhaohui Qian[1,2], Weixiang Ye[4], Yongshun Zeng[1,2], Xiaoxing Peng[3]

and Xianwu Luo[1,2,†]

[1]Beijing Key Laboratory of $CO_2$ Utilization and Reduction Technology, Department of Energy and Power Engineering, Tsinghua University, Beijing 100084, China;

[2]State Key Laboratory of Hydro-science and Engineering, Tsinghua University, Beijing 100084, China;

[3]National Key Lab on Ship Vibration and Noise, China Ship Scientific Research Center, Wuxi 214082, China;

[4]Institute of Nuclear and New Energy Technology, Tsinghua University, Beijing, 100084, China;

† Email address for correspondence: luoxw@tsinghua.edu.cn

**Abstract:**

The discrete tone radiated from a cavitating tip vortex, known as 'vortex singing', was first recognized in 1989, but its sound generation mechanism has remained a mystery for over thirty years. In this letter, by means of the correction for the cavitation bubble dynamics and the dispersion relation of cavity interfacial waves, we found that after the far-end disturbances propagate upstream, the whistling vortex should be triggered by near-end sound sources, the breathing mode waves. Further utilizing the theoretical solutions for singing lines and the potential singing cavitation number with frequency, we accurately identified all available tests for seeking the vortex singing over the past three decades, answering a long-standing perplexity: why such a best whistler is able to appear only within a narrow range of the cavitation number.

## 1. Introduction

Vortex sound is a fundamental phenomenon in nature and has long been accepted as a basic model in fluid mechanics (Howe 2003). Once the cavitation occurs under the lower pressure, the vortex-induced sound pressure intensifies (Brennen 2014).

Specifically, cavitation has been an attractive topic in hydraulic machinery and ocean engineering for over one hundred years (Arndt *et al.* 1991; Luo *et al.* 2022). Among various types of cavitation, the tip vortex cavitation (TVC) is the first one to appear on propellers or axial-flow turbines, acting as a crucial source for underwater noise and vibrations (Posa *et al*. 2022; Qian *et al*. 2022; Ji *et al.* 2023). Consequently, many studies focused on forecasting TVC's inception (Arndt & Keller 1992; Amini *et al.* 2019; Chen *et al.* 2019), capturing general features such as the vortex trajectory, vapor core radius and roll-up kinematics (Asnaghi *et al.* 2020; Xie *et al.* 2021; Cheng *et al.* 2021; Russell *et al.* 2023), with few discussions concerning on the cavity dynamics and sound generation mechanisms (Pennings *et al.* 2016a; Liu & Wang 2019; Klpwijk *et al.* 2022; Wang *et al.* 2023).

The phenomenon of 'vortex singing', a notable noise enhancement associated with the interfacial instability of TVC, was initially documented by Higuchi (1989) and later termed by Maines & Arndt (1997). They tested different hydrofoils by varying the velocity, angle of attack and water quality in two facilities, one in *Obernach*, Germany and the other at St. Anthony Falls Laboratory (*SAFL*), USA. The whistling vortex was observed within a narrow range of cavitation number, typically leading to a notable peak of 20 dB to 30 dB above the sound spectrum. Briançon-Marjollet and Merle (1997) also reported observations of the singing vortex over a different hydrofoil in *G.T.H.*, a larger cavitation tunnel. Experiments conducted by Pennings *et al.* (2015) at *Delft* University of Technology did not hear the vortex singing, although the recorded TVC dynamics exhibited similarities to that at *SAFL* and *Obernach*. Next attempts were conducted at China Ship Scientific Research Center (*CSSRC*) by Peng *et al.* (2017) and Song *et al.* (2018), they regenerated the vortex singing and noted the significant impact of inflow air content ($D_O$) on both the singing cavitation number and the desinent cavitation number. Recent studies at Zhejiang University (*ZJU*) by Ye *et al.* (2021, 2023) further revealed the propagation of strong breathing waves along TVC, however, the vortex singing was not detected. Jiang *et al.* (2024) found more intriguing conditions where the singing is stimulated simultaneously with the disappearance of the vortex cavity.

So far, there is still not much understanding regarding how the vortex singing generates and why the vortex singing can only occur within a very narrow range of cavitation number (Arndt 2002). Traditional hydrodynamic research approaches seem inadequate for such a real-scale cavitating phenomena. The prospects for experimental and numerical simulation studies to interpret this whistle also appear less optimistic (Klapwijk *et al.* 2022; Simanto *et al.* 2023). To emphasize, several approaches have been utilized to describe the singing waves through the resonance frequency of two-dimensional (2-D) axisymmetric cavitation bubbles (Choi & Ceccio 2007; Choi *et al.* 2009; Bosschers 2018a) and the three-dimensional (3-D) cavity interfacial dynamics defined by a semi-empirical corrected dispersion relationship (Pennings *et al.* 2015; Pennings 2016b; Bosschers 2018b). However, due to the lack of profound physical insights, these attempts ultimately ended in failure. Overall, predicting the singing cavitating vortex is important and challenging, especially for the complex vortex-cavitation-sound interplays.

Therefore, in this letter, we aim to comprehensively discuss the sound generation mechanism of the singing vortex from a theoretical perspective, expecting to unveil the mysterious veil of this whistler.

## 2. Theory

### 2.1 Resonance frequency of a 2-D cavitation bubble

According to Franc & Michel (2016), the dynamic equation describing the radial

($r$) motion for the wall of a 2-D axisymmetric cavitation bubble, $R(t)$, is given as,

$$\frac{1}{r}(R\ddot{R}+\dot{R}^2-\frac{1}{r^2}R^2\dot{R}^2-u_\theta^2)=-\frac{1}{\rho_L}\frac{\partial p}{\partial r} \qquad (2\text{-}1)$$

here $\rho_L$ and $p$ respectively represent the liquid density and pressure. Assuming that the distribution of liquid tangential velocity, $u_\theta$, satisfies the viscous *Lamb-Oseen* vortex model, further neglecting the surface tension and non-condensable gas nuclei, so that Equation (2-1) can be integrated from $R(t)$ to $r_\infty$ (usually replaced by the half-width of the cavitation tunnel), and consequently the linearization yields a standard second-order oscillation system, which is written as,

$$\ddot{K}+2\xi\omega_n\dot{K}+\omega_n^2 K=0 \qquad (2\text{-}2)$$

$K(t)$ is the radial perturbation of $R(t)$ relative to the cavitating vortex core radius, $r_{cv}$, and $\xi$ denotes the damping ratio. The angular frequency $\omega_n = 2\pi f_n$, where $f_n$ represents the natural oscillation frequency of the bubble,

$$f_n=\frac{U_\infty}{2\pi r_c}\sqrt{\sigma K_\sigma(\sigma/\sigma_d)\left/\ln\frac{r_\infty}{r_c}\right.} \qquad (2\text{-}3)$$

$r_c$ is the equilibrium radius (replaced by the mean radius of TVC), and the cavitation number is $\sigma = (p_\infty - p_v)/(0.5\rho_L U_\infty^2)$, where $p_v$ is the saturation vapor pressure. It should be emphasized that $\sigma_d$ is the theoretical desinent cavitation number, and $K_\sigma$ indicates a stiffness term,

$$\begin{cases} K_\sigma = \dfrac{1+e^{-2ak_c^2}-2e^{-ak_c^2}}{1+e^{-2ak_c^2}-2e^{-ak_c^2}+2ak_c^2 Ei(-2ak_c^2)-2ak_c^2 Ei(-ak_c^2)} \\ \dfrac{1+e^{-2ak_c^2}-2e^{-ak_c^2}+2ak_c^2 Ei(-2ak_c^2)-2ak_c^2 Ei(-ak_c^2)}{2a\ln 2k_c^2}=\dfrac{\sigma}{\sigma_d} \end{cases} \qquad (2\text{-}4)$$

which is calculated by the intermediate parameter $k_c$, that is, the ratio of cavity core radius to the vortex core radius in equilibrium state, and it is finally derived that $K_\sigma$ is only related to the relative cavitation number $\sigma/\sigma_d$.

**2.2 Dispersion relations for 3-D cavity interfacial waves**

Due to the fact that a 2-D theory can only predict the oscillation frequency but cannot explain why such a whistle is triggered only within a narrow range of $\sigma$, it is more appropriate to utilize the 3-D cavity interfacial dispersion relation to analyze the vibration modes for a singing TVC, further determine the frequency and wavelength of whistling waves. According to Bosschers *et al.* (2018b), for small-scale axial phase velocities or low-frequency oscillations, the dimensionless dispersion relation is,

$$\tilde{\omega}^\pm(\kappa,k_\theta)=\frac{2\pi f^\pm r_c}{U_\infty}=\frac{U_c}{U_\infty}\kappa+\frac{V_c}{U_\infty}\left[k_\theta\pm\sqrt{\frac{-|\kappa|K'_{k_\theta}(|\kappa|)}{K_{k_\theta}(|\kappa|)}}\right] \qquad (2\text{-}5)$$

where $\tilde{\omega}$ denotes the 3-D resonant frequency, $\kappa = k_x r_c \cong -ik_r r_c$. The radial wavenumber,

$k_r$, is always positive however the axial wavenumber, $k_z$, may have positive as well as negative values, and $k_\theta$ refers to the azimuthal wavenumber. $U_c$ and $V_c$ respectively represent the mean axial and azimuthal velocity along the TVC surface, $K_{k_\theta}$ refers to a modified *Bessel* function of the second kind (Bosschers 2008). The plus and minus sign in the right-hand-side of Equation (2-5) corresponds to two frequencies of each cavity vibration mode. Furthermore, $k_\theta = 0$, $\pm 1$ and $\pm 2$ respectively correspond to the monopole breathing mode, dipole bending mode wave, and quadrupole double helical mode wave (Pennings *et al.* 2015).

The key to reproducing TVC interfacial dynamics lies in defining the coefficients $U_c/U_\infty$ and $V_c/U_\infty$ reasonably. Using the potential flow assumption and viscous *Lamb-Oseen* vortex model, $V_c/U_\infty$ is calculated as $\sigma^{0.5}$ and $(\sigma K_\sigma)^{0.5}$, respectively. Moreover, due to the lack of a suitable analytical model for evaluating $U_c$, it is typically treated as $U_c/U_\infty = 1.0$. However, we can also run the *Bernoulli* law based on the tangential viscosity-correction and deduce $U_c/U_\infty = (\sigma+1-\sigma K_\sigma)^{0.5}$. According to these discussions, three coefficient correction models, labeled as *I* to *III*, are listed in table 1.

**TABLE 1.** The velocity ratio parameters used in Equation (2-5) determined by model *I* to *III*.

| Model | *I* | *II* | *III* |
|---|---|---|---|
| $V_c/U_\infty$ | $\sigma^{0.5}$ | $(\sigma K_\sigma)^{0.5}$ | $(\sigma K_\sigma)^{0.5}$ |
| $U_c/U_\infty$ | 1.0 | 1.0 | $(\sigma+1-\sigma K_\sigma)^{0.5}$ |

### 2.3 Sound sources and predictions

The interfacial waves travelling on a TVC are regarded as sound sources, and the solution for the amplitude of acoustic pressure perturbation, $p^*$, radiated by a vibrating cylindrical vortex cavity with infinite length is found as (Bosschers 2018b),

$$p^*(r) = \rho_L \omega_{eff}^2 r^* \frac{K_{k_\theta}(|k_x r|)}{\left|k_x K'_{k_\theta}(|k_x r_c|)\right|} \tag{2-6}$$

here $r^*$ corresponds to the amplitude of each cavity vibration mode, and the effective angular frequency $\omega_{eff}$ is defined as,

$$\omega_{eff}^2 = \left(U_\infty k_x + \frac{\Gamma_\infty k_\theta}{2\pi r^2} - \omega\right)\left(U_c k_x + \frac{V_c k_\theta}{r_c} - \omega\right) \tag{2-7}$$

where $\Gamma_\infty$ denotes the vortex strength. The sound pressure level (SPL) is calculated by $20\log_{10}(p^*/p_{ref})$, and the underwater reference pressure, $p_{ref}$, is $1\mu Pa$.

## 3. Results

### 3.1. A perfect whistler

The vortex singing experiments conducted at *CSSRC* are utilized to validate the theories given in **Theory** section. Under conditions of $U_\infty = 7m/s$ and $D_O = 68\%$, the measured $\sigma_d$ is around 2.30 (Song *et al.* 2017). And prior to singing, the TVC's surface typically exhibits a stationary twisted shape. Subsequently, by slowly increasing the inflow cavitation number from $\sigma = 1.25$ to $\sigma = 1.40$, a sharp whistle emerged, and then the pressure was stabilized to keep the singing wave for several minutes. As shown in figure 1(a), the high-speed photography captures a dancing TVC from the *xy* plane view, and the time-space distribution law of cavity radius is calculated using *Canny*'s edge detection method. In figure 1(b), the far-field sound pressure level (SPL) spectrum has been plotted for both singing ($\sigma$ = 1.40) and non-singing ($\sigma$ = 1.45) conditions. The hydrophone is positioned 112.5*mm* above the tip of the hydrofoil, revealing the SPL increase to 154*dB* at the singing frequency $f_s$ = 320*Hz*, with similar increments at the harmonics of $2f_s$ and $3f_s$. Further observation of figure 1(a) indicates rhythmic oscillations of the singing cavity radius over time at $f_s$, with twisted cavity surface generating a shuttle-shaped structure on the spatiotemporal diagram. And the wavelength of the helical mode wave is $\lambda_{exp} \approx 23.5mm$. Averaging the radius between $x/C$ = 0.5 and 1.56 yields $r_c$ = 1.30*mm*, where the error is one pixel (0.18*mm*). The convection line for $U_\infty$ (white dashed line) is also depicted and shows an opposite slope with the oscillating stripes in the background, suggesting that the disturbance waves is propagating upstream instead. Therefore, the speculation that the vortex singing source is the standing wave over TVC (Maines & Arndt 1997), is probably a misconception that requires clarification.

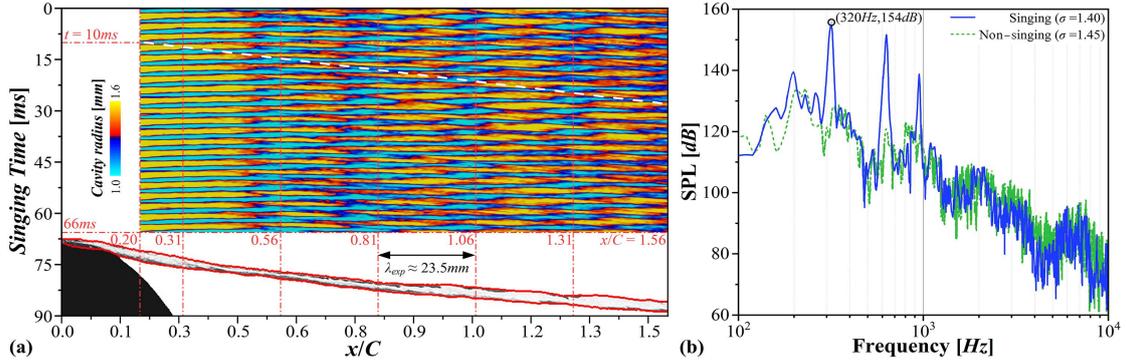

**FIGURE 1.** (a) Variation of 'dancing' cavity radius in time and space from the *xy* plane view at the singing condition ($\sigma_s$ = 1.40, $\sigma_d \approx$ 2.30) provided by *CSSRC*, with the (b) SPL spectrum recorded by the hydrophone for singing and non-singing conditions.

To analyze the mechanism of cavity interfacial wave emitting the whistle, a 2-D Fast Fourier Transform (FFT) was performed on the varying TVC radius, resulting in dimensionless frequency-wavenumber grayscale plots in figure 2 as the backgrounds. Apart from the mean mode wave ($\tilde{\omega}$ = 0, $\kappa$ = 0), the brightest point that represents the highest amplitude is located at $f_s$ = 320*Hz*, furthermore, two inclined bright stripes are observed. Figures 2(a) to (c) respectively illustrate the theoretical dispersion curves predicted using models *I* to *III* given by table 1 presented in **Theory** section. Here the zero-frequency points attached with $k_\theta$ = -2$^+$ represent the stationary wave of the double helical mode, denoted as ($\tilde{\omega}_{zf}, \kappa_{zf}$), and zero-group-velocity ($\partial\tilde{\omega}/\partial\kappa$ = 0) points with $k_\theta$ = 0$^-$ represent the traveling wave of the breathing mode, characterized by ($\tilde{\omega}_{zgv}, \kappa_{zgv}$). Clearly, in figure 2(a), none of the curves predicted with the potential flow assumption align with the high-amplitude region on the background. After improving from the model *I* to model *II*, the zero-group-velocity point is pulled closer to $f_s$, with the curves of double helical modes ($k_\theta$ = 2$^-$ and $k_\theta$ = -2$^+$) approaching the inclined bright stripes in the first and fourth quadrant. For model *III* depicted in figure 2(c), the helical mode curves further match with the bright stripes, indicating that the axial velocity correction enhances the accuracy of wavelength predictions and brings ($\tilde{\omega}_{zgv}$, $\kappa_{zgv}$) closer to the bright region around $f_s$. Additionally, the intersection point of helical mode curves with the zero-frequency line coincides with the second high-amplitude bright point, representing the stationary double helical wave. Its wavelength is predicted as $\lambda_{zf} = \kappa_{zf}/r_c = 24.4mm$, close to the singing wavelength, $\lambda_{exp}$ (i.e. 23.5*mm*). Therefore, model *III* demonstrates a satisfactory accuracy for predicting TVC's interfacial singing dynamics.

Since the zero-group-velocity point in figure 2(c) locates near the intersection point of the two dispersion curves, $k_\theta = -2^+$ and $k_\theta = 0^-$, it is still confusing whether the oscillation comes from the breathing wave or the double helical wave. To accurately identify the dominant acoustic source, the following radius reconstruction formula is utilized for visualizing three potential cavity vibration modes with highest amplitude in figure 2(c),

$$\begin{cases} R_{zgv\_0^-}(x,\theta,t) = r_c + A_{zgv} \exp\left[i\left(-2\pi x/\lambda_{zgv} + 2\pi f_{zgv} t\right)\right] \\ R_{zgv\_-2^+}(x,\theta,t) = r_c + A_{zgv} \exp\left[i\left(-2\pi x/\lambda_{zgv} - 2\theta + 2\pi f_{zgv} t\right)\right] \\ R_{zf}(x,\theta,t) = r_c + A_{zf} \exp\left[i\left(-2\pi x/\lambda_{zf} - 2\theta + 2\pi f_{zf} t\right)\right] \end{cases} \quad (3\text{-}1)$$

where the parameters such as amplitude, frequency, and wavelength used in the above equation are determined by theoretical model *III* or experiments, as table 2 exhibits. The frequency of zero-group-velocity point on the curve of $k_\theta = 0^-$, $f_{zgv}$, is calculated as 324*Hz*, which is very close to the singing frequency, $f_s$ (i.e. 320*Hz*). The theoretical wavelength, $\lambda_{zgv} = 151.3mm$, is more than six times longer than that of the stationary double helical wave, $\lambda_{zf} = 24.4mm$. Moreover, in the second quadrant of the frequency-wavenumber spectrum, the phase velocity of the zero-group-velocity point is predicted as, $V_p = \tilde{\omega}_{zgv}/\kappa_{zgv} = -49.0m/s$, representing a breathing wave shooting from the downstream to the tip. The zero-group-velocity wave for the breathing mode ($R_{zgv\_0^-}$) and the double helical mode ($R_{zgv\_-2^+}$), together with the zero-frequency wave for the double helical mode ($R_{zf\_-2^+}$) are respectively visualized in figure 2(d).

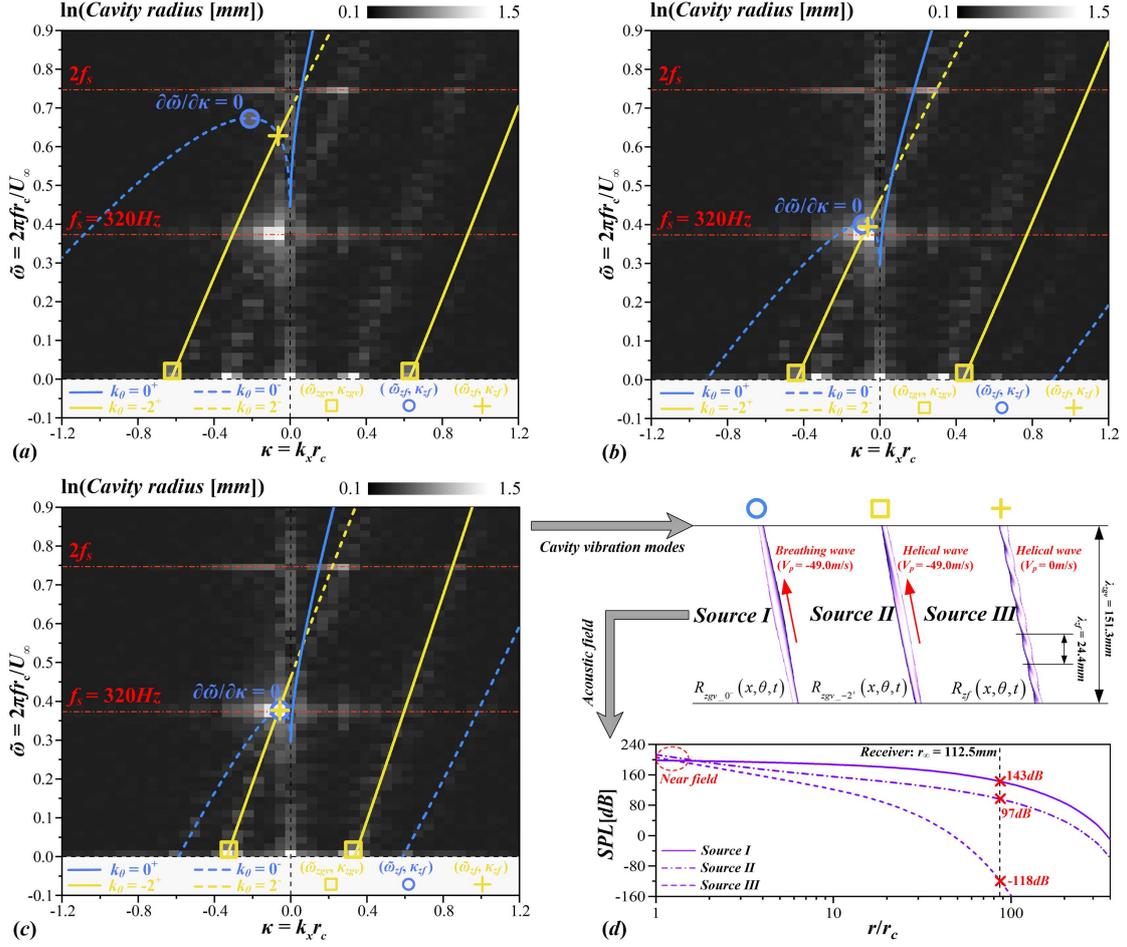

**FIGURE 2.** Frequency-wavenumber diagrams of the variation of cavity radius on the $xy$ plane at the singing condition ($\sigma_s$ = 1.40, $\sigma_d \approx$ 2.30) from *CSSRC*, the dispersion coefficients for comparison are obtained by (a) Model *I*, (b) Model *II* and (c) Model *III*. (d) The theoretical sound pressure level (SPL) for three types of the near-field sound sources ($R_{zgv\_0^-}$, $R_{zgv\_-2^+}$ and $R_{zf}$) solved by model *III*.

**TABLE 2.** The parameters applied for reconstructing TVC's interfacial waves at the singing condition conducted in *CSSRC* ($\sigma_s$ = 1.40, $\sigma_d \approx$ 2.30).

| Parameters | $A_{zf}$ [mm] | $f_{zf}$ [Hz] | $\lambda_{zf}$ [mm] | $A_{zgv}$ [mm] | $f_{zgv}$ [Hz] | $\lambda_{zgv}$ [mm] |
|---|---|---|---|---|---|---|
| Value | 0.21 | 0 | 24.4 | 0.36 | 324 | 151.3 |

Finally, using Equation (2-6) in **Theory** section, distributions of sound pressure level (SPL) for three types of interfacial wave sources (*I* to *III*) are derived and plotted in figure 2(d). Near the bubble's wall, the sound pressure is mainly contributed by two helical mode waves. However, in the far-field where the hydrophone is located ($r_\infty$ = 112.5$mm$), the source of $R_{zgv\_0^-}$ maintains a leading advantage of 30 to 40$dB$ over that of $R_{zgv\_-2^+}$, and the SPL for $R_{zf\_-2^+}$ rapidly decays. This indicates that during the vortex singing, the zero-group-velocity point on $k_\theta$ = 0$^-$ are sufficiently excited to conduct vapor volume pulsations, thereby causing a significant increase in the far-field sound pressure for a discrete tone by tens of decibels. Due to the lack of theoretical and experimental techniques, earlier studies often mistakenly identify this traveling wave of breathing mode as the standing wave along the TVC. We now have the reason to infer that TVC is becoming a perfect whistler only when the zero-group velocity wave of breathing mode is correctly triggered.

### 3.2. The singing mechanism

Based on above discussions, the singing waves can be simplified and reproduced as $R_s(x, \theta, t)$ using the mean wave, the zero-group-velocity wave ($k_\theta = 0^-$) which dominates the sound field as well as the zero-frequency wave ($k_\theta = -2^+$) that affects the cavity shape,

$$R_s(x,\theta,t) = r_c + R_{zgv\_0^-} + R_{zf} \tag{3-2}$$

In figure 3(b), four typical instants are selected, the singing interfacial waves are depicted based on Equation (3-2), which are in good agreement with the experimental snapshots. The breathing waves are predicted to generate from the downstream with the phase velocity of $V_p = -49.0 m/s$ and propagate towards the tip, continuously being absorbed by the boundary layer separation bubbles at the tip. This means that the peak of the breathing wave (*BP*) near $x/C = 1.22$ at $T_0$ should retrogress to around $x/C = 0.53$ at $T_1 = T_0 +1.33ms$. Similarly, the valley of breathing wave (*BV*) near $x/C = 0.78$ at $T_2$ should theoretically be transported to around $x/C = 0.09$ at $T_3$. These predictions are all consistent with real variations of TVC in *CSSRC*'s experiment, furthermore, the predicted wavelength of the stationary double helical wave, $\lambda_{theor} = \lambda_{zf} = 24.4mm$, is also close to the experimental measurement, $\lambda_{exp} = 23.5mm$. Therefore, once the cavity starts to sing, the double helical waves remain basically stationary to shape the TVC, while the breathing waves continuously propagate from downstream to the tip, contributing to the sound production.

Considering the collapse stage at the far end of the TVC, where the axial flow has fully developed, therefore the TVC can be approximated as the 2-D axisymmetric bubble. Then prediction based on the linearized bubble dynamics and Equation (2-3) in **Theory** section gives $f_n = 319Hz$, which is sufficiently close to $f_s$. In figure 3(a), the dynamic process of the bubble's wall, $R(t)$, tangential velocity, $u_\theta(t)$, and liquid pressure, $p(t)$, during the step response of a 2-D cavitation bubble at the far end is further illustrated. From $T_1$ to $T_2$, the tangential velocity is significant, leading to a decrease in liquid pressure around the bubble according to the conservation of angular momentum, resulting in the bubble growth. Conversely, from $T_2$ to $T_3$, as $u_\theta$ decreases, $p$ increases, driving the bubble collapse. Such instabilities at the far end of TVC can be triggered by self-excited oscillations, however, they are often insufficient to keep the vortex singing. Therefore, it is more effective to adjust the far-field cavitation number to induce a step change in pressure, aiming to stimulate disturbances with large amplitude and certain frequency at the far end.

Overall, when the far-end TVC is excited by the step change in far-field pressure, initial disturbances will be generated. Subsequently, these disturbances are moving upstream along the helical surface of the TVC in the form of breathing wave, which plays the role of dominant acoustic source. The simplified interfacial waves perfectly interpret the sound generation mechanism of this dancing cavity.

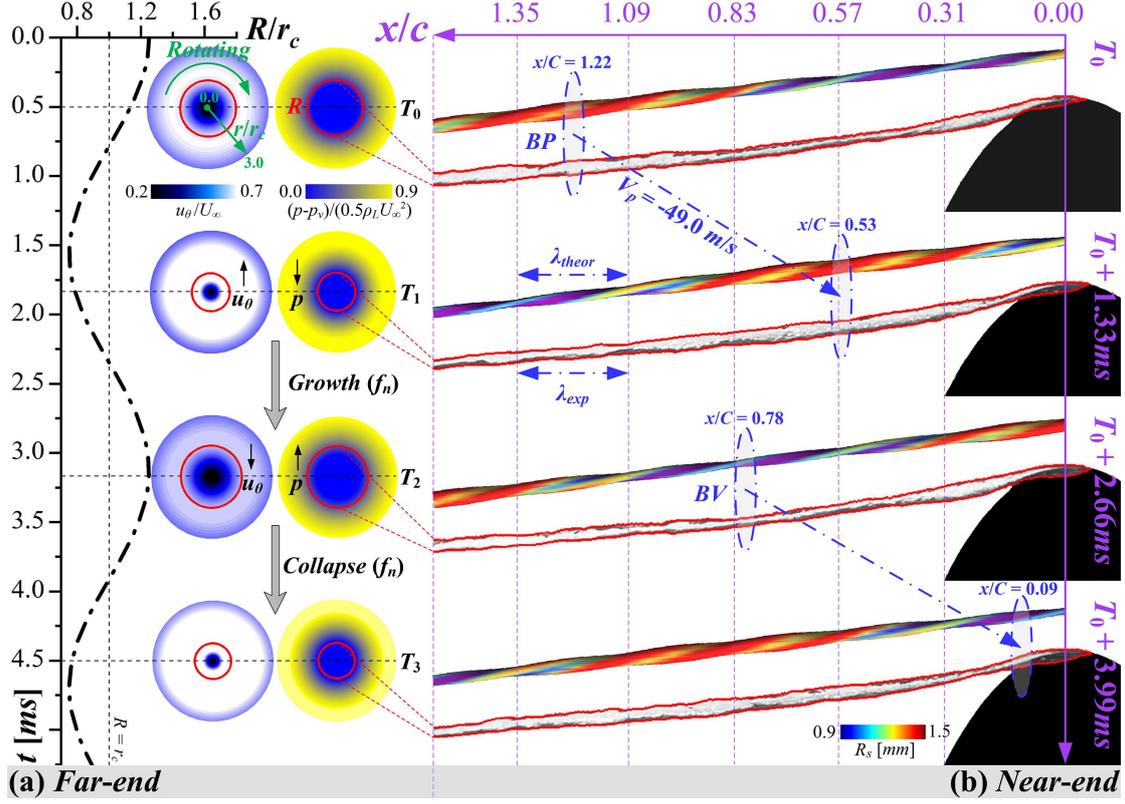

**FIGURE 3.** Four typical instants within a singing cycle from the side view of *CSSRC*'s tunnel, where (a) the 2-D cavitation bubble dynamics at the far-end TVC are illustrated with radius $R(t)$, tangential velocity $u_\theta(t)$ and liquid pressure $p(t)$, (b) the whistling waves at the near-end TVC are simplified using Equation (3-2) and parameters in table 2 for the singing condition ($\sigma_s = 1.40$, $\sigma_d \approx 2.30$).

### 3.3. Narrow singing tones

#### 3.3.1. Three types of singing lines

Although the preceding discussion is limited to one singing condition in *CSSRC*, it offers a fresh perspective on understanding the triggering mechanism of a whistling vortex. The further contemplation becomes, how to accurately and sufficiently excite the zero-group-velocity waves at $k_\theta = 0^-$ after the far-end disturbance being uploaded to the stationary helical surface in the upstream. This is exactly the key to the success or failure of vortex singing. Therefore, in this chapter, we should propose a necessary condition to discern whether TVC can enhance the narrow tones.

Here, a natural inference from figure 3(c) is that if $f_i \sim \kappa_i$, i.e., the intersection of $f_0^- \sim \kappa$ and $f_{-2}^+ \sim \kappa$ is sufficiently approaching $f_{zgv} \sim \kappa_{zgv}$, where $df_0^- / d\kappa = 0$ is located, then the zero-group-velocity wave will be accurately triggered and further sustained. Therefore, $f_i = f_{zgv}$ becomes one necessary condition. Similarly, when $f_j \sim \kappa_j$, i.e., the intersection frequency of $f_0^+ \sim \kappa$ and $f_2^- \sim \kappa$ is also sufficiently close to $f_{zgv} \sim \kappa_{zgv}$, then the zero-group-velocity wave should also be excited. Hence, (1) $f_i = f_{zgv}$ and (2) $f_j = f_{zgv}$ together serve as the analytical criteria for determining vortex singing. According to **Theory** section, the 3-D resonance frequencies of TVC are calculated as,

$$\begin{cases} \tilde{\omega}^-(\kappa,0) = \dfrac{2\pi f_{0^-} r_c}{U_\infty} = \kappa\sqrt{\sigma+1-\sigma K_\sigma} - \sqrt{\sigma K_\sigma}\sqrt{\dfrac{-|\kappa|K'_0(|\kappa|)}{K_0(|\kappa|)}} \\[2mm] \tilde{\omega}^+(\kappa,0) = \dfrac{2\pi f_{0^+} r_c}{U_\infty} = \kappa\sqrt{\sigma+1-\sigma K_\sigma} + \sqrt{\sigma K_\sigma}\sqrt{\dfrac{-|\kappa|K'_0(|\kappa|)}{K_0(|\kappa|)}} \\[2mm] \tilde{\omega}^+(\kappa,-2) = \dfrac{2\pi f_{-2^+} r_c}{U_\infty} = \kappa\sqrt{\sigma+1-\sigma K_\sigma} + \sqrt{\sigma K_\sigma}\left(-2+\sqrt{\dfrac{-|\kappa|K'_{-2}(|\kappa|)}{K_{-2}(|\kappa|)}}\right) \\[2mm] \tilde{\omega}^-(\kappa,2) = \dfrac{2\pi f_{2^-} r_c}{U_\infty} = \kappa\sqrt{\sigma+1-\sigma K_\sigma} + \sqrt{\sigma K_\sigma}\left(2-\sqrt{\dfrac{-|\kappa|K'_2(|\kappa|)}{K_2(|\kappa|)}}\right) \end{cases} \quad (3\text{-}3)$$

By two criteria of $\tilde{\omega}^-(\kappa, 0) = \tilde{\omega}^+(\kappa, -2)$ and $\tilde{\omega}^+(\kappa, 0) = \tilde{\omega}^-(\kappa, 2)$, the dimensionless wavenumbers, $|\kappa_i| = 0.062$ and $|\kappa_j| = 0.062$, are solved at the intersections. Therefore, the criteria (1) and (2) can be respectively expressed by $\tilde{\omega}^+(\kappa_i, -2) = \tilde{\omega}^-(\kappa_{zgv}, 0)$ and $\tilde{\omega}^-(\kappa_j, 2) = \tilde{\omega}^-(\kappa_{zgv}, 0)$, yielding the following relationship,

$$1 + 1/\sigma = \gamma K_\sigma(\sigma/\sigma_d) \quad (3\text{-}4)$$

where the criterion (1) includes two sets of solutions, namely (a) $\gamma = 3.45$ and (c) $\gamma = 247.26$, and (b) $\gamma = 1.42$ corresponds to the only solution for criterion (2). Apparently, the necessary conditions for triggering a whistling TVC is that there must be a determined relationship between the local cavitation number ($\sigma$) and the cavitation desinent point ($\sigma_d$). As $\sigma_d$ changes, these relations are transformed into three curves of $\tilde{\omega} = 2\pi f r_c/U_\infty$ that is varying with $\sigma^{0.5}$,

$$\begin{cases} \text{Singing-line-}a, & \tilde{\omega}(\sigma) = 0.263\sqrt{\sigma+1} \\ \text{Singing-line-}b, & \tilde{\omega}(\sigma) = 0.524\sqrt{\sigma+1} \\ \text{Singing-line-}c, & \tilde{\omega}(\sigma) = 0.025\sqrt{\sigma+1} \end{cases} \quad (3\text{-}5)$$

As shown in figure 4, theoretical singing lines (a) and (b) pass exactly through the singing conditions provided by *CSSRC* and *G.T.H.*, which firstly validates the correctness of our theoretical criteria. From the Equation (3-4), it is directly deduced that as $\sigma/\sigma_d \to 0$, $K_\sigma \to 1$, which indicates that there are three limit cavitation numbers (0.41, 2.37 and 0.004) at the left of three singing lines, respectively.

Next, the physical meaning of each singing line is discussed using the maximum desinent cavitation number ($\sigma_d = 15.70$) predicted from *G.T.H.*, under this condition, there are four intersections generated by $f_{zgv} \sim \sigma^{0.5}$ and $f_i \sim \sigma^{0.5}$. Among them, only when $\sigma$ is very small or $\sigma/\sigma_d$ is very close to 1, two hidden intersection points appear. And the trajectory of these two points with respect to $\sigma_d$ forms the black singing frequency line (c) in figure 4. And this line indicates that there always exists a vortex singing interval near $\sigma_d$, which also explains a series of dynamic singing conditions near the cavitation desinent point that are recently discovered by Jiang *et al.* (2024). Because of the gentle slope of curve (c), the fact that singing frequencies tested in their results are all concentrated in a narrow range is understandable. Excluding the intersection points at both ends, the trajectory line of the other two intersections between $f_{zgv} \sim \sigma^{0.5}$ and $f_i \sim \sigma^{0.5}$ generates the singing frequency line (a). Due to the moderate cavitation number and frequency range, this line becomes a preferred theoretical solution for previous vortex singing experiments. Finally, the singing line (b) extended by two intersection points of $f_{zgv} \sim \sigma^{0.5}$ and $f_j \sim \sigma^{0.5}$. Noting that this line only appears when $\sigma_d$ is relatively large, and it better depicts the frequency distribution for *G.T.H.*'s singing conditions. Importantly, the appearance of several unexpectedly vortex singing conditions in this region used to be confusing in earlier studies.

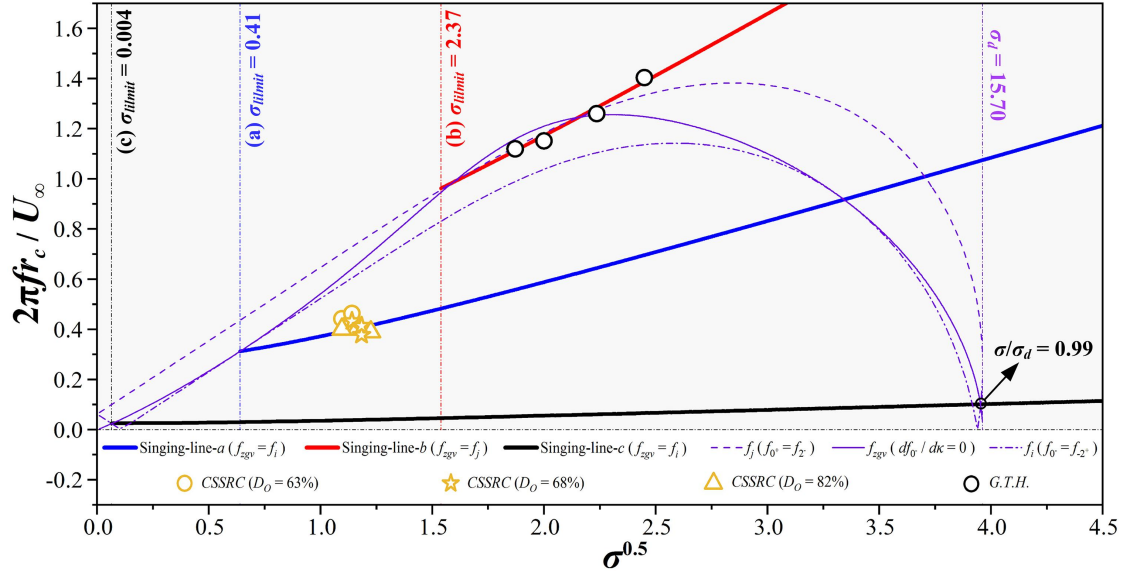

**FIGURE 4.** Theoretical solutions for the frequency of three vortex singing lines varying with $\sigma^{0.5}$, with singing conditions from *CSSRC* and *G.T.H.*, further limited by cavitation number at (a) 0.41, (b) 2.31 and (c) 0.004.

The underlying idea employed in determining three types of theoretical singing lines is the same, which is to confine the intersection frequency range of helical waves and breathing waves to ensure they precisely fall near the zero-group-velocity points. This enables the sustained triggering of the breathing waves on the surface of TVC, resulting in whistling. After addressing the trigger mechanism and the acoustic source, we are left with one final question unanswered: why is vortex singing triggered only within a narrow range of the cavitation number?

### 3.3.2. Sing or not to sing

For over three decades, this question has remained unresolved. A comprehensive solution requires, firstly, accurate prediction of the cavitation numbers and frequency ranges associated with all vortex singing conditions at *SAFL*, *Obernach*, *CSSRC* and *G.T.H.*, secondly, convincing explanations need to be provided for the inability to hear the whistling vortex in test conditions at facilities such as *Delft* and *ZJU*. To ensure the consistency in our response and further generalize our theory for available singing or non-singing conditions, the dimensionless process should be performed with the maximum cavity radius, $r_{max}$. The relation between $r_{max}$ and $r_c$ is deduced from the overshoot for a second order bubble oscillating system, $M_p$,

$$0 \ (\zeta \to 1) \leq M_p = e^{-\zeta\pi/\sqrt{1-\zeta^2}} = (r_{max} - r_c)/r_c \leq 1 \ (\zeta \to 0) \tag{3-6}$$

Once the damping coefficient, $\zeta$, approaches zero, the maximum radius reaches twice the equilibrium radius, further combining $r_{max}/r_c \geq 1$, the new dimensionless vortex singing frequency, $\tilde{\omega}_s = 2\pi f_s r_{max}/U_\infty$, is constrained by,

$$2\pi f_s r_c/U_\infty \leq \tilde{\omega}_s \leq 4\pi f_s r_c/U_\infty \tag{3-7}$$

and the natural frequency, $\tilde{\omega}_n = 2\pi f_n r_{max}\ln^{0.5}(r_\infty/r_{max})/U_\infty$, is limited by,

$$2\pi f_n r_c \sqrt{\ln(r_\infty/r_c)}/U_\infty \leq \tilde{\omega}_n \leq 4\pi f_n r_c \sqrt{\ln(r_\infty/r_c)}/U_\infty \tag{3-8}$$

hence, the upper and lower limit lines of frequency are determined by $r_{max}/r_c = 2$ and $r_{max}/r_c = 1$, respectively representing the maximum and minimum vibration amplitude of TVC. The triggering theory can only be validated if the operating points fall within the narrow ranges predicted for both $\tilde{\omega}_s$ and $\tilde{\omega}_n$.

To predict the potential $\sigma$ ranges for vortex singing with the cavitation desinent number corresponding to each experimental facility, a slight relaxation of the criterion $f_i = f_{zgv}$ proposed in chapter 3.3.1 is conducted,

$$\left| \frac{f_i(\sqrt{\sigma}, \sigma_d) - f_{zgv}(\sqrt{\sigma}, \sigma_d)}{f_{zgv}(\sqrt{\sigma}, \sigma_d)} \right| \leq 1.0\% \tag{3-9}$$

where the choice of a 1% error band is to strike a balance between the reliability and accuracy. The criterion (3-9) is employed to predict the vortex singing or non-singing conditions at *ZJU*, *Delft*, *SAFL* and *Obernach* successively in figures 5(a) to (f). Since the measurements at *CSSRC* have been validated in chapters 3.1 and 3.2, while *G.T.H.* has not provided required dimensional information about the singing TVC, they are not discussed in this chapter.

In figure 5(a), one condition provided by *ZJU* is depicted, where the pronounced breathing cavity was observed but no vortex singing could be heard. Utilizing the measured $\sigma_d = 11.90$ (Ye *et al.* 2023), we have plotted curves of $f_i$, $f_{zgv}$, $2f_i$ and $2f_{zgv}$ and the purple singing band satisfying the 1% error criterion on the $\tilde{\omega}_s \sim \sigma^{0.5}$ spectrum. Theoretically, there are four singing bands corresponding to four intersection points generated by $f_{zgv}$ and $f_i$. However, according to chapter 3.3.1, we only need to focus on the conditions on the singing line (a), hence the singing bands when $\sigma$ is very small or $\sigma/\sigma_d$ is very close to 1 have been removed. It is observed that the experimental point (3.34, 0.57) does not fall within any of the theoretical singing bands, $\sigma = 0.3 \sim 0.7$ and $7.2 \sim 9.4$, but instead land in the white region on the right. This indicates that the breathing wave triggered at *ZJU* is insufficient and unstable, prone to switching to the double helical wave, which further induces the disappearance of the whistle.

For the other test at *ZJU*, the measured $\sigma_d$ decreases to 7.63 (Ye *et al.* 2021), yet only the pronounced breathing cavity surface was captured without vortex singing, as shown in figure 5(b). Two singing bands are located by the criterion (3-9), with the position of the left band almost unchanged, however the $\sigma$ range of the right singing band is reduced overall to $\sigma$ = 4.4~5.9. Nonetheless, the test point (2.62, 0.54) still falls within the white area on the far right, indicating that the zero-group-velocity wave has still not been adequately excited, resulting in strong TVC oscillations but no sound production.

Next are the conditions at *Delft*, marked by blue circles in figure 5(c). Although Bosschers (2018a) have accurately replicated the dynamics of TVC interfacial waves under these conditions by a semi-empirical corrected dispersion relation, they failed to explain why only weak breathing wave were observed in their tests without forming discrete peaks in the SPL spectrum. Utilizing the estimated $\sigma_d$ = 3.77 provided by Pennings *et al.* (2015), we calculated the theoretical singing bands by criterion (3-9), resulting in $\sigma$ = 0.3~0.7 and 1.3~2.8. Further analysis based on $f_i = f_{zgv}$ reveals the most probable singing locations to be $\sigma_{ml}$ = 0.4 and $\sigma_{mr}$ = 2.4. As figure 5(c) shows, the experimental points at *Delft* fall within these two cavitation numbers, with frequencies consistently below the lower limit line of the singing bands. Therefore, the zero-group-velocity wave has not been triggered, and the vortex singing is hard to achieve relying on weak breathing waves.

Thus far, a clear explanation for the inability to hear the vortex whistling in *Delft* and *ZJU* has been presented. Following this, a series of singing conditions found by *SAFL* and *Obernach* should be discussed. Since most early experiments did not record the cavitation desinent points, and these singing conditions are mostly corresponding to the singing line (a) given in chapter 3.3.1, thus we can estimate $\sigma_d$ by calculating the mean cavitation number of each test based on Equation (3-4).

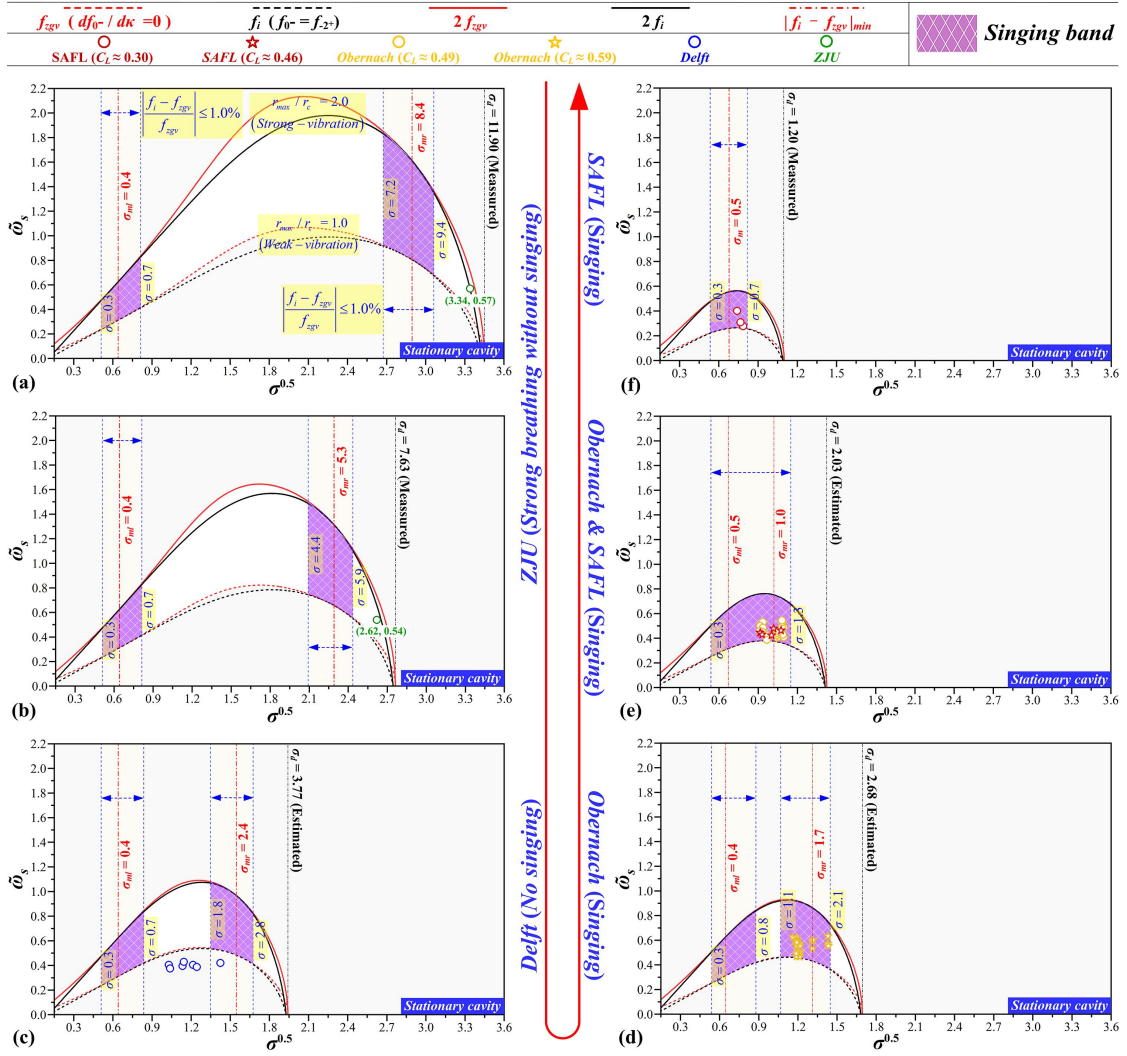

**FIGURE 5.** Predicted singing bands limited by $|(f_i - f_{zgv})/f_{zgv}| \leq 1.0\%$ at different test tunnels. (a) *ZJU*, $\sigma_d = 11.90$, strong breathing without singing, (b) *ZJU*, $\sigma_d = 7.63$, strong breathing without singing, (c) *Delft*, $\sigma_d = 3.77$, weak breathing but no singing, (d) *Obernach*, $\sigma_d = 2.68$, singing, (e) *Obernach & SAFL*, $\sigma_d = 2.03$, singing, (f) *SAFL*, $\sigma_d = 1.20$, singing.

As figure 5(d) shows, in a series of singing tests conducted at *Obernach*, the lift coefficient is $C_L \approx 0.59$, the mean cavitation number is $\sigma = 1.56$, hence the estimated $\sigma_d = 2.68$. Subsequently, the analytic left and right singing bands fall within the ranges of $\sigma = 0.3 \sim 0.8$ and $\sigma = 1.1 \sim 2.1$, respectively. Compared to figures 5(a)~(c), there is a significant increase in the proportion of vortex singing bands, with a tendency for the left and right bands to move closer together. Further, all singing points land in the right side of the singing band, and are closer to the lower limit frequency line, which indicates that zero-group-velocity waves are sufficiently excited and going to whistle at specific frequencies.

As for *Obernach* ($C_L \approx 0.49$) and *SAFL* ($C_L \approx 0.46$), $\sigma_d$ is estimated as 2.03, as shown in figure 5(e). According to criterion (3-9), the theoretical left and right singing bands start to merge into one part, located by $\sigma = 0.3 \sim 1.3$. By determining the most probable singing conditions based on $f_i = f_{zgv}$, $\sigma_{ml} = 0.5$ and $\sigma_{mr} = 1.0$ are deduced.

Moreover, all tested singing points fall within the purple singing band, being closer to $\sigma_{mr}$ and the lower frequency limit line. Therefore, the singing vortex is successfully heard in experiments.

Finally, for singing conditions at *SAFL* ($C_L \approx 0.30$), the measured minimum desinent cavitation number is, $\sigma_d = 1.20$. As figure 5(f) shows, the theoretical solution reveals that the singing band is located at $\sigma = 0.3 \sim 0.7$, with the proportion of purple singing band decreasing. Although there is no solution for $f_i = f_{zgv}$, the most probable singing condition can be determined through $| f_i = f_{zgv} |_{min}$, which is found to be $\sigma_m = 0.5$. At this point, all singing points fall within the singing band, located at the right of $\sigma_m$. Hence the zero-group-velocity waves are correctly triggered.

### 3.3.3. Fully triggered, under-triggered or untriggered

In chapter 3.3.2, we have predicted the presence of vortex singing within what cavitation number and frequency range once the TVC starts oscillating from its static helical regime. Combining with conclusions in chapter 3.2, we have realized that the natural oscillation at the far-end breakdown segment of the cavitating vortex rope will determine the oscillating frequency for the entire TVC. This also relates to whether the zero-group-velocity waves at the near end of TVC can be accurately triggered. To further study the triggering conditions of the far-end bubble oscillations, we plotted the theoretical $f_n$, $2f_n$ and the blue triggering band on the $\tilde{\omega}_n \sim \sigma^{0.5}$ diagram in figure 6. To be emphasized, due to the limited data from early experiments, only several points with determinable parameters are selected for analysis in this chapter.

As shown in figures 6(a) to (b), in two sets of tests at *ZJU*, the dimensionless natural oscillation frequency, $\tilde{\omega}_n$, falls within the white frequency band on the right. This indicates that although the far-end TVC can oscillate at the expected frequency, it fails to continuously trigger the breathing waves for the near end of TVC, and belongs to an 'under-triggered' state. In figure 6(c), all measured points at *Delft* locate below the lower frequency limit line, indicating that the far-end TVC cannot be stimulated. Therefore, zero-group-velocity waves are suppressed, placing TVC in an 'untriggered' state. As shown in figure 6(d), all vortex singing conditions in *Obernach* ($C_L \approx 0.59$) concentrate within the upper and lower bounds, located on the right of the theoretical triggering frequency interval. Therefore, after the far-end TVC starts to oscillate and transfer disturbance upstream, at the near end, the breathing waves can be excited at the expected frequency, placing TVC in a 'fully triggered' state. For *Obernach* ($C_L \approx 0.49$) and *SAFL* ($C_L \approx 0.46$), as figure 6(e) illustrates, all vortex singing points are distributed between the upper and lower bounds of the triggering interval, near $\sigma_{mr} = 1.0$. Therefore, the singing waves are correctly triggered and further heard. Finally, in figure 6(f), the singing points from *SAFL* ($C_L \approx 0.30$) also land in the blue triggering interval, successfully triggering the whistling TVC.

In this chapter, by the criterion of $| (f_i - f_{zgv})/f_{zgv} | \leq 1.0\%$, we have addressed a long-standing question: why the vortex singing, reported over the past three decades, occurs only within a narrow range of cavitation number. By $\sigma_d$ and $\sigma/\sigma_d$, the singing band and triggering interval are determined, helping us gain a profound understanding of the mechanisms and conditions of both the far-end oscillation and the near-end sound generation along TVC. Further, it holds the potential to guide experimental and industrial design to effectively control and utilize the unstable and acoustic flows associated with the vortex cavitation.

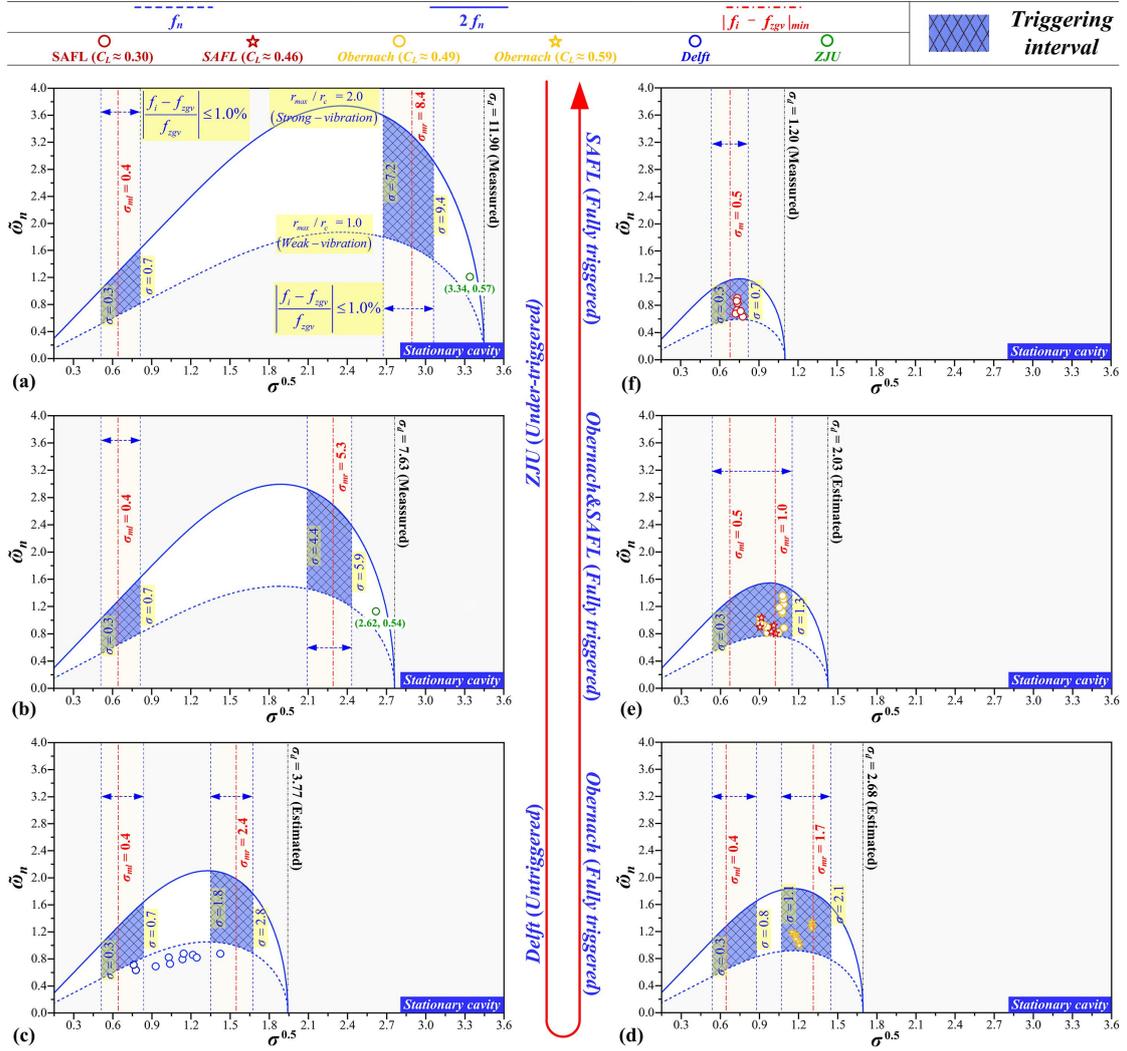

**FIGURE 6.** Predicted triggering intervals limited by $|(f_i - f_{zgv})/f_{zgv}| \leq 1.0\%$ at different test tunnels. (a) *ZJU*, $\sigma_d = 11.90$, under-triggered, (b) *ZJU*, $\sigma_d = 7.63$, under-triggered, (c) *Delft*, $\sigma_d = 3.77$, untriggered, (d) *Obernach*, $\sigma_d = 2.68$, fully triggered, (e) *Obernach & SAFL*, $\sigma_d = 2.03$, fully triggered, (f) *SAFL*, $\sigma_d = 1.20$, fully triggered.

## 4. Conclusions

The vortex singing, a typical noise enhancement induced by the severe interfacial instability over a tip vortex cavity, has been recognized as one of the most challenging topics in cavitation hydrodynamics. In this letter, the theoretical solutions provide us a comprehensive understanding of such a best whistler,

(1) Based on the improved 2-D oscillation frequency and 3-D interfacial dispersion relations, the vortex singing is verified to be triggered by the far-end cavity oscillation and generated by the zero-group-velocity breathing waves along the near-end surface. The cavitation number ($\sigma$) and desinent cavitation number ($\sigma_d$) help to determine the acoustic source and the featured oscillation frequency ($f_i, f_j, f_{zgv}$ and $f_n$).

(2) Three analytical singing lines are derived by the condition of $f_i = f_{zgv}$ and $f_j = f_{zgv}$, the first two are confirmed by singing conditions in *CSSRC* and *G.T.H*, besides, the third line explains the phenomenon of TVC's disappearing while singing near $\sigma_d$ that has been recently reported.

(3) By employing the criterion of $| (f_i - f_{zgv})/f_{zgv} | \leq 1.0\%$, the theoretical singing band (bounded by $f_i$ and $f_{zgv}$) and triggering interval (bounded by $f_n$) are located. Concerning conditions in *SAFL*, *Obernach*, *G.T.H.* and *CSSRC*, the far-end cavity oscillations can adequately stimulate $f_{zgv}$, whereas at *Delft* and *ZJU*, $f_{zgv}$ remains under-triggered or untriggered. These findings not only demonstrate when and how the vortex sings or refuses to sing, but also response to a longstanding question about why the cavitating vortex only whistles in a narrow range of $\sigma$.

# Acknowledgements

This work was supported by the National Natural Science Foundation of China (No: 52336001), the China Postdoctoral Science Foundation (No: 2023M731895), and the International Partnership Program of Chinese Academy of Sciences (No: 025GJHZ2022118FN). The authors would also like to acknowledge professor Y. Q. Liu from Shanghai Jiao Tong University and professor Q. Q. Ye from Zhejiang University for their contributions in the analytical study.

# Competing interests

The authors declare no competing interests.

# Data availability

The test datasets that we used for validation are publicly available. All data supporting the findings in this study are available within the article. Additional data are available from the corresponding authors upon reasonable request.